\documentclass[apj]{emulateapj}

\usepackage{psfig}
\usepackage{wasysym}
\usepackage{natbib}
\newcommand{\msini}{\ensuremath{M \sin{i}}}
\newcommand{\feh}{\ensuremath{[\mbox{Fe}/\mbox{H}]}}
\newcommand{\vth}{$^{\mbox{th}}$}

\newcommand{\persec}{\ensuremath{\mbox{s}^{-1}}}

\newcommand{\mjup}{\ensuremath{\mbox{M}_{\mbox{Jup}}}}

\newcommand{\Mjupmath}{M_{\mbox{Jup}}}
\def\astrosun {\mbox{$\odot$}}
\newcommand{\Msol}{\ensuremath{\mbox{M}_{\astrosun}}}
\shorttitle{Possible 2:1 Resonance for HD 37124}
\shortauthors{Wright et al.}

\begin{document}

\title{The California Planet Survey III.  A Possible 2:1 Resonance in
  the Exoplanetary Triple System HD 37124\altaffilmark{1}}
\author{J. T. Wright\altaffilmark{2,3}} 
\author{Dimitri Veras\altaffilmark{4}, Eric B. Ford\altaffilmark{4}}
\author{John Asher Johnson\altaffilmark{5}}
\author{G. W. Marcy\altaffilmark{6}, A. W. Howard\altaffilmark{6,7},
  H. Isaacson\altaffilmark{6}} 
\author{D. A. Fischer, J. Spronck\altaffilmark{8}} 
\author{J. Anderson\altaffilmark{9}, J. Valenti\altaffilmark{9}}

\altaffiltext{1}{Based on observations obtained
at the W. M. Keck Observatory, which is operated jointly by the
University of California and the California Institute of Technology.
The Keck Observatory was made possible by the generous financial
support of the W. M. Keck Foundation.}
\altaffiltext{2}{Department of Astronomy, 525 Davey Lab, The Pennsylvania State
  University, University Park, PA 16802} 
\altaffiltext{3}{Center for Exoplanets and Habitable Worlds, The Pennsylvania State
 University, University Park, PA 16802}
\altaffiltext{4}{Department of Astronomy, University of Florida, 211 Bryant Space Science Center, P.O. Box 112055,  Gainesville, FL 32611-2055}
\altaffiltext{5}{Department of Astronomy, California Institute of
  Technology, MC 249-17, Pasadena, CA}
\altaffiltext{6}{Department of Astronomy, 601 Campbell Hall, University of California, Berkeley, CA 94720-3411}
\altaffiltext{7}{Townes Postdoctoral Fellow, Space Sciences
  Laboratory, University of California, Berkeley}
\altaffiltext{8}{Astronomy Department, Yale University, New Haven, CT}
\altaffiltext{9}{Space Telescope Science Institute, 3700 San Martin
  Dr., Baltimore, MD 21218}

\begin{abstract}
We present new radial velocities from Keck Observatory and
both Newtonian and Keplerian solutions for the triple-planet system
orbiting HD 37124. The orbital solution for the system has
improved dramatically since the third planet was first reported in \citet{Vogt05} with
an ambiguous orbital period.  We have resolved this ambiguity, and
show that the outer two planets have an apparent period
commensurability of 2:1.  A dynamical analysis finds both resonant and
non-resonant configurations consistent with the radial velocity data, and constrains the mutual
inclinations of the planets to be $< \sim 30^\circ$.  

We discuss HD 37124 in the context of the other 19 exoplanetary systems with
apparent period commenserabilities, which we summarize in a table.
We show that roughly one in three well-characterized multiplanet systems has a
apparent low-order period commensuribility, which is more than would
na\"ively be expected if the periods of exoplanets in known
multiplanet systems were drawn randomly from the observed distribution of
planetary orbital periods.  

\end{abstract}

\section{Introduction}
\label{Intro}

To date, over 50 exoplanetary systems with more than one planet have
been discovered, including: the extraordinary detections of the first exoplanets
orbiting the pulsar PSR B1257+12 \citep{Wolszczan92,Wolszczan94}; the imaged system
orbiting HR 8799; those discovered during the microlensing event
OGLE-2006-BLG-109L \citep{Gaudi08}; several systems discovered by
transit, including four or five\footnote{KOI 877 may be a blend of
  two, separately transiting systems.} {\it multiply} transiting systems from the Kepler
mission \citep{Steffen10}; and 43 systems discovered by radial
velocity (RV) searches \citep{Wright09d}.  The RV systems include the four-planet systems
$\mu$ Ara \citep{Santos04b,Pepe07}, GJ 581 \citep{Mayor09} and GJ 876
\citep{Rivera05,Rivera10} and the five-planet system orbiting 55
Cancri \citep{Fischer08}.   Of all these multplanet systems, only four
are known to host three or more giant\footnote{$\msini > 0.2 \Mjupmath$} planets with well-determined
orbital paramaters:  $\upsilon$ And \citep{Butler_upsand}, HIP 14810
\citep{Wright09c}, $\mu$ Ara \citep{Pepe07} and HD 37124
\citep{Vogt05}.   

HD 37124 (HIP 26381) is a  0.85 \Msol\ metal-poor
\citep[\feh=-0.44,][]{SPOCS} G4 dwarf (V=7.7).
\cite{Vogt00} announced the a Jovian, $P \sim 150$ d planet
orbiting HD 37124 from HIRES data taken at Keck Observatory as part of
the California and Carnegie Planet Search.  Further monitoring of the
star revealed substantial long-term residuals.    \cite{Butler03} fit
these residuals with an eccentric, 1940 d planet, but noted that the
solution was not unique (and \cite{Gozdziewski03} showed
  that this fit was, in fact, unstable.)   

After collecting two more years of data, \citet{Vogt05} was able to
report the detection of a third planet in the system, though with an
ambiguity:  while the $b$ and $c$ components had  
clearly defined periods, the $d$ component could be fit nearly equally
well with periods of either 2300 d or 29.32 d, the latter likely
being an alias due to the lunar cycle.\footnote{Time on the Keck
  telescopes dedicated to observing 
  bright, planet search targets with HIRES is usually assigned during
  bright or gray time;  the resulting scarcity of data points during
  new moon can interact with planetary signals to create
  spurious, aliased solutions.}   \citet{Wright09d} reported that
recent Keck velocities had resolved the ambiguity qualitatively in
favor of the longer orbital period.  \citet{Gozdziewski06b} explored
the many possible dynamical configurations consistent with the
\citet{Vogt05} velocities, including many resonant solutions.
\citet{Gozdziewski08} used the system to demonstrate a fast MENGO
algorithm, but they did not explore the 2:1 resonance, as the data did
not seem to favor it at the time.

We present new Keck observations, and these data provide a unique orbital solution for the outer planet.
The outer planet period we find is more consistent with the
original period reported by \citet{Butler03} than the refined orbit of \citet{Vogt05}\footnote{\citet{Vogt05}
  opted to refer to the new, 840 d signal as the $c$ component, despite the
prior 1940 d fit of \citet{Butler03}, because that prior fit was
so speculative, and because their new fit put the very existence of a
1940 d periodicity in some doubt.}  (though we find a much lower
eccentricity).   Herein, we present the entire history of
Keck velocities obtained for this star, and present self-consistent orbital solutions showing that
the outer two planets are in or very near a 2:1 mean-motion
resonance (MMR).  This is the 20th exoplanetary system to be found near an
MMR, and  only the tenth system with an apparent 2:1 commensurability.

Period commensurabilities (PCs) represent important dynamical
indicators in the Solar System and have been linked with observables and formation
mechanisms \citep{Goldreich65}.  The near-5:2 PC of Jupiter and Saturn,
also known as ``The Great Inequality", might be the remnant of a
divergent resonant crossing that produced the current architecture of the outer
Solar System, the Late Heavy Bombardment, and the Trojan Asteroids
\citep{Gomes05,Morbidelli05,Tsiganis05,Tsiganis05b}.  The populations
of the asteroid belt and the Kuiper Belt, exemplified by  
the PC and near-PC-populated Kirkwood Gaps \citep[e.g.][]{Tsiganis02}
the Plutinos (3:2 PCs with Pluto and Neptune)  and the twotinos \citep[2:1 PCs with Pluto and Neptune;
e.g. ][]{Murray-Clay05,Chiang02b}, have 
implications for the migratory history of Jupiter and Neptune and the prospect of, e.g. secular
resonant sweeping \citep[e.g., ][]{Nagasawa08}.  Near-PCs found in satellite and ring systems
have had direct observational consequences; the Saturnian satellite Pandora was $\sim 19^{\circ}$
behind its predicted orbital longitude in a 1995 ring plane crossing \citep{French03} due to its
121:118 PC with neighboring satellite Prometheus.   

By extension, we may anticipate similar importance in the 
growing number of exoplanetary systems exhibiting PCs.  In extrasolar
systems, Mean Motion Resonances (MMRs) have been interpreted as the
indication of convergent migration in multi-planet systems,
\citep[e.g.][]{Thommes03,KleyResonance,Papaloizou05}. Several subsequent studies \citep{Beauge06,  Terquem07,
  Pierens08,Podlewska08,Podlewska09,Libert09,Rein09,
  Papaloizou10,Rein10,Zhang10b,Zhang10c} exploring convergent migration for a
variety of masses, separations and disk properties have found many regions of mass and
orbital element phase space in which planets are easily captured
through this mechanism.  

\section{Velocities and Orbital Solution}

Table~\ref{velocities} contains radial velocity measurements for HD
37124 from the HIRES spectrograph  \citep{Vogt94} at Keck Observatory
obtained by the California Planet Survey consortium using the iodine
technique \citep{Butler96b}.  Note that the quoted errors are our internal (random) errors, 
with no ``jitter'' included \citep{Wright05}. 

These velocities supersede our previously published  
velocities for this star, as we continue to refine our data reduction
pipeline.   Our ever-evolving radial velocity pipeline is descended in
spirit and form from that described in \citet{Butler96b}, but includes many small and large
technical improvements, a thorough discussion of which is beyond the
scope of this manuscript.  Some details can be found in \S~4.1 of
\citet{Howard10a}, \S~3 of \citet{Howard09}, and in Batalha et al.\
2011 (ApJ, accepted).  

On issue of instant relevance is that in August 2004 the HIRES CCD
detector was upgraded to a CCD mosaic. 
The old Tektronix 2048 EB2 engineering-grade CCD displayed
a variable instrumental profile asymmetry due to a charge transfer inefficiency which manifested itself as 
small changes in a star's measured radial velocity as a function of exposure
time (i.e. raw counts on the chip.)  We
apply an emperical, spectral-type dependent model to correct this effect for velocities measured
prior to the detector change.  The new CCD mosaic shows no evidence of this effect, but
as a consequence of the switch there is a small velocity offset between data sets that span
the two detector sets similar to the detector-to-detector
offsets discussed in \citet{Gregory10}.  These offsets could, in
principle, be different for every target.  

Analysis of RV standards and known planetary systems show that such an
offset is usually small -- of order 5 m/s -- and very often consistent
with zero.   As a result, we report two independent data sets for this system in
Table~\ref{velocities}, one from each of the two detectors.  We solve
for the detector offset as an unconstrained free parameter.  The times of
observation are given in JD-2440000.

\begin{figure*}
\caption{Radial velocity curves for the HD 37124 triple system.\label{fig}}
\plotone{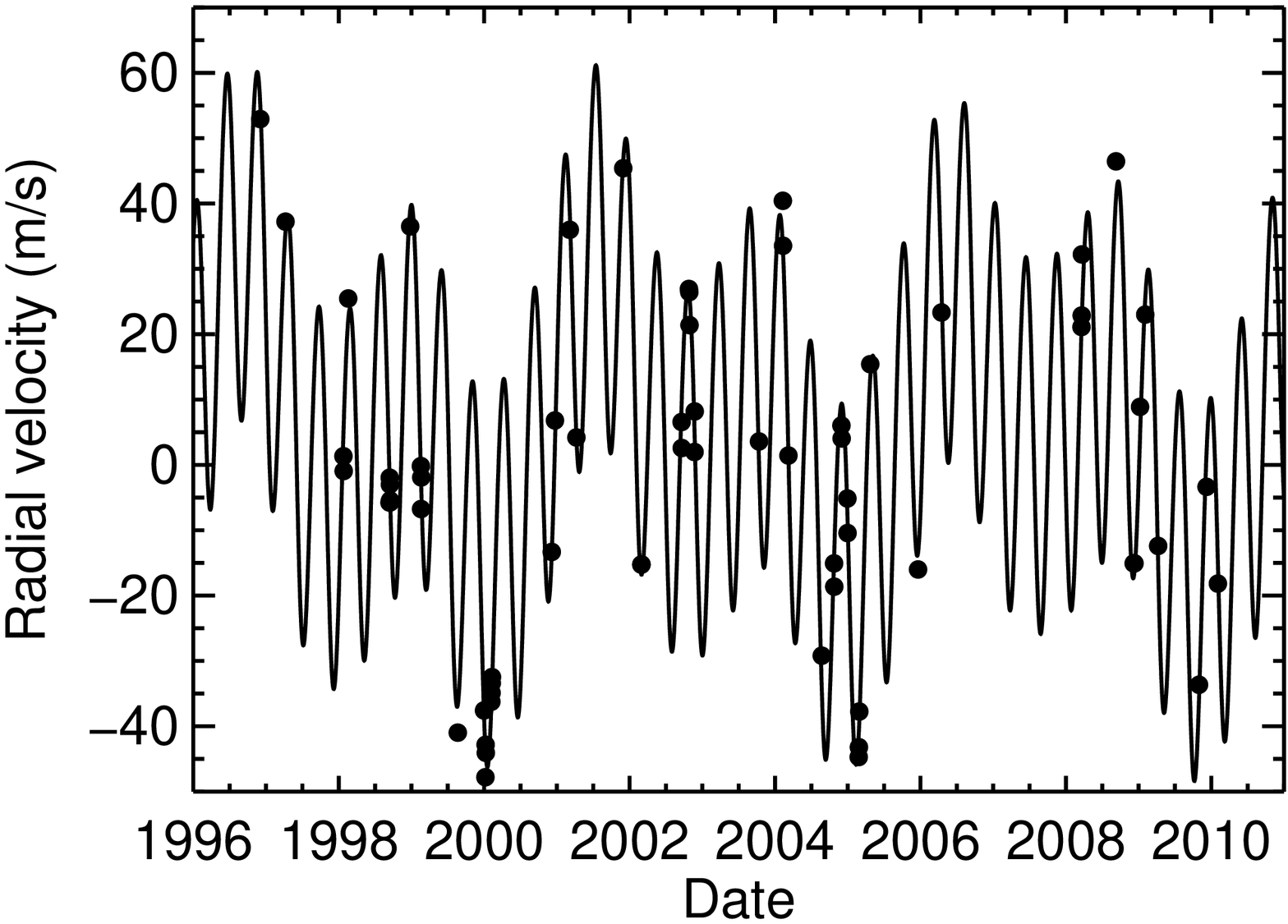}
\plotone{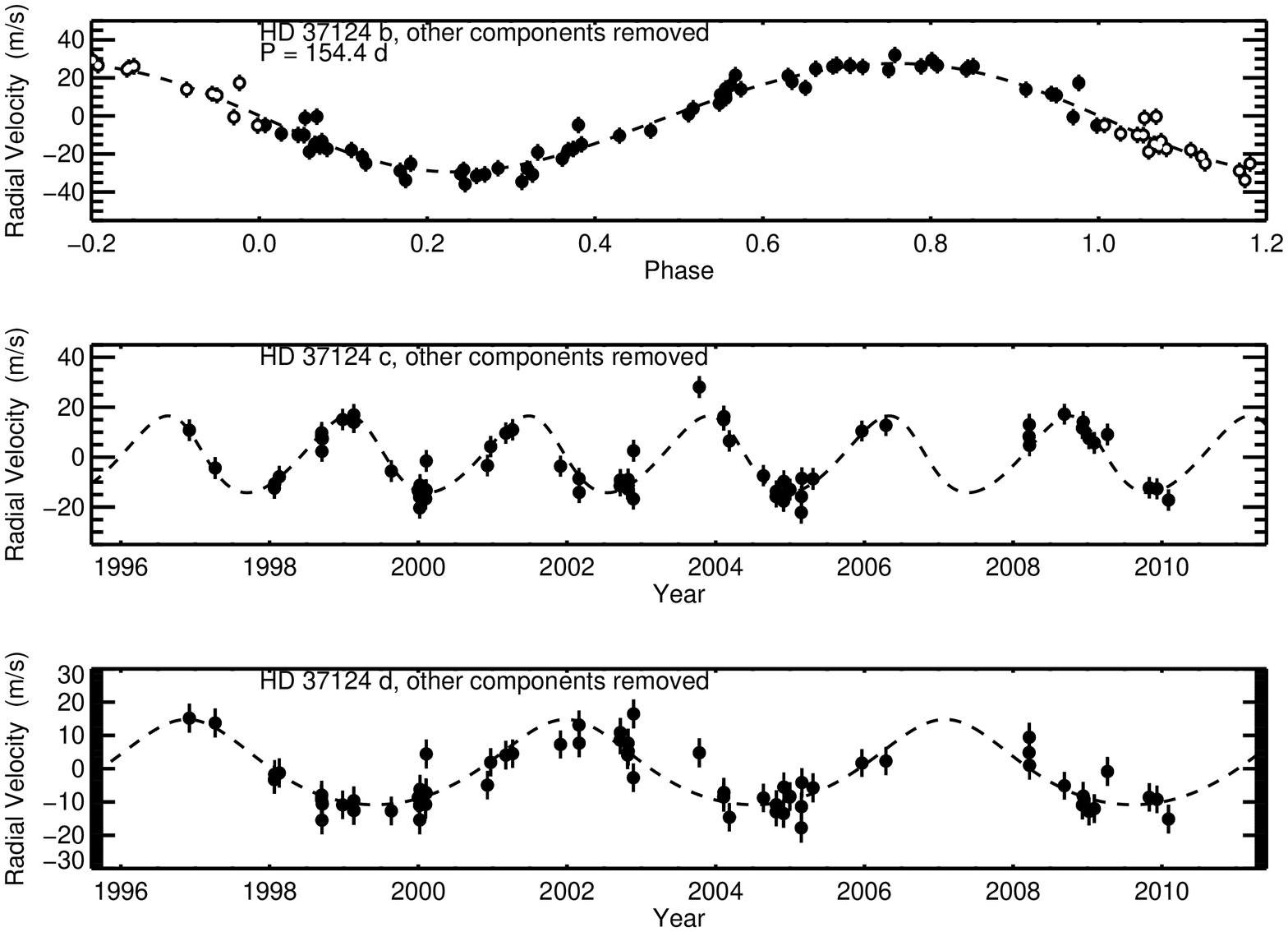}
\end{figure*}

We fitted the data using the publicly available multi-planet
RV-fitting IDL package \texttt{RV\_FIT\_MP}, described in
\citet{Wright09b}.  In Table~\ref{orbital} we  present our
3-planet Keplerian (kinematic) fit,\footnote{This solution is of similar quality to the best fit
Newtonian solution, and is dynamically stable.  We consider it representative of the
ensemble of good Newtonian solutions.} which yields 
r.m.s. residuals of 4.4 m\persec, and we plot the fit and velocities
in Figure~\ref{fig}.  We find a best-fit offset between CCDs of $4.8$
m\persec.  The orbital parameters and their uncertainties were
determined from 100 bootstrapped trials \citep[as described
in][]{Marcy05,Butler06,Wright07}.  The orbital fits and dynamical analysis herein
are put forth under the assumption that the velocities are not
detectably influenced by additional, unmodeled planets in the system.   We have
integrated these orbital parameters for 10 Myr using the methods
described in \S\ref{Newton} assuming coplanarity, and found them to
yield a stable configuration.  

\begin{deluxetable}{cccc}
\tablecolumns{3} 
\tablecaption{Radial Velocities for HD 37124 \label{velocities}}
\tablehead{\colhead{Time} & \colhead{Velocity} & \colhead{Uncertainty}
  & CCD \\
 \colhead{JD-2440000} &
  \colhead{m \persec} & \colhead{m \persec} & }
\startdata 
10420.04655 & 54.81 & 1.5 & 1 \\
10546.73646 & 28.74 & 1.2 & 1 \\
10837.76625 & 6.32 & 1.6 & 1 \\
10838.94868 & 6.38 & 1.7 & 1 \\
10861.80464 & 17.89 & 1.4 & 1 \\
11069.03617 & -3.99 & 1.4 & 1 \\
11070.13190 & -1.66 & 1.3 & 1 \\
11071.11494 & 1.28 & 1.6 & 1 \\
11072.12947 & -11.58 & 1.5 & 1 \\
11073.02962 & -8.72 & 1.3 & 1 \\
11172.89571 & 39.40 & 1.6 & 1 \\
11226.78065 & -0.48 & 1.4 & 1 \\
11227.78167 & -2.05 & 1.4 & 1 \\
11228.74293 & -11.01 & 1.3 & 1 \\
11412.14161 & -33.76 & 1.6 & 1 \\
11543.98278 & -31.36 & 1.4 & 1 \\
11550.94262 & -44.57 & 1.4 & 1 \\
11551.94008 & -46.20 & 1.5 & 1 \\
11552.89162 & -47.97 & 1.5 & 1 \\
11580.76121 & -36.28 & 1.8 & 1 \\
11581.83559 & -36.17 & 1.4 & 1 \\
11582.78849 & -37.36 & 1.4 & 1 \\
11583.72387 & -35.69 & 1.5 & 1 \\
11884.04436 & -14.73 & 1.6 & 1 \\
11900.03518 & 3.01 & 1.4 & 1 \\
11974.80019 & 39.56 & 1.4 & 1 \\
12007.74522 & 4.48 & 1.5 & 1 \\
12242.99064 & 48.20 & 1.5 & 1 \\
12333.94545 & -19.13 & 1.7 & 1 \\
12334.78556 & -12.82 & 1.7 & 1 \\
12536.12848 & 10.64 & 1.8 & 1 \\
12537.08597 & 10.28 & 1.7 & 1 \\
12573.03767 & 27.38 & 1.6 & 1 \\
12574.99934 & 27.48 & 1.7 & 1 \\
12576.02212 & 23.58 & 1.6 & 1 \\
12600.99996 & 8.06 & 1.7 & 1 \\
12602.03213 & 7.02 & 1.7 & 1 \\
12925.01639 & 13.03 & 1.7 & 1 \\
13044.77359 & 36.56 & 1.6 & 1 \\
13045.74638 & 32.78 & 1.5 & 1 \\
13072.85948 & -2.85 & 1.7 & 1 \\
13240.13983 & -34.52 & 1.5 & 2 \\
13302.13193 & -4.28 & 1.6 & 2 \\
13302.97959 & -5.69 & 1.4 & 2 \\
13338.96446 & 15.77 & 1.2 & 2 \\
13340.09520 & 16.45 & 1.5 & 2 \\
13368.88930 & -6.51 & 1.0 & 2 \\
13369.78156 & -6.60 & 1.0 & 2 \\
13425.87197 & -39.98 & 1.4 & 2 \\
13426.82980 & -39.12 & 1.3 & 2 \\
13428.77030 & -36.85 & 1.3 & 2 \\
13483.72749 & 19.49 & 1.0 & 2 \\
13723.90630 & -7.28 & 1.6 & 2 \\
13841.76427 & 34.09 & 1.4 & 2 \\
14544.83023 & 25.40 & 1.6 & 2 \\
14545.78169 & 26.99 & 1.4 & 2 \\                                                      
14546.78977 & 24.28 & 1.3 & 2 \\
14718.08322 & 43.60 & 1.6 & 2 \\
14806.91704 & -7.74 & 1.5 & 2 \\
14810.89031 & -9.50 & 1.7 & 2 \\
14838.94681 & 0.39 & 1.8 & 2 \\
14864.95362 & 29.45 & 1.8 & 2 \\
14929.76349 & -5.49 & 1.7 & 2 \\
15135.00085 & -21.29 & 1.6 & 2 \\
15172.99171 & 12.60 & 1.6 & 2 \\
15229.78574 & -12.32 & 1.6 & 2 \\
\enddata
\end{deluxetable}

\begin{deluxetable}{cccc}
\tablecolumns{3}
\tablecaption{Best-Fit Kinematic Orbital Elements for Exoplanets in the
 HD 37124 System\label{orbital}}
\tablehead{\colhead {Parameter} & \colhead{b} & \colhead{c} & \colhead {d}}
\startdata

$P$ (d)                     &       154.378 $\pm$ 0.089 & 885.5 $\pm$ 5.1 & 1862 $\pm$ 38 \\          
$T_p$ (JD-2440000) &                10305 $\pm$ 11 & 9534 $\pm$ 11 & 8558 $\pm$ 11 \\                 
$e$                           &     0.054 $\pm$ 0.028 & 0.125 $\pm$ 0.055 & 0.16 $\pm$ 0.14 \\         
$\omega$ (degrees)  &               130\tablenotemark{a}& 53 $\pm$ 17 & 0\tablenotemark{a} \\         
$K$ (m\persec)         &            28.50 $\pm$ 0.78 & 15.4 $\pm$ 1.2 & 12.8 $\pm$ 1.3 \\             
\msini  (\mjup)           &         0.675 $\pm$ 0.017 & 0.652 $\pm$ 0.052 & 0.696 $\pm$ 0.059 \\      
a (AU)                       &      0.53364 $\pm$ 0.00020 & 1.7100 $\pm$ 0.0065 & 2.807 $\pm$ 0.038 \\
\hline
R.M.S.                       & \multicolumn{3}{c}{4.03}    \\
jitter                         &     \multicolumn{3}{c}{4 \mbox{m/s}} \\
$\chi^2_\nu$            & \multicolumn{3}{c}{0.8}\\
$N_{\mbox{obs}}$         &      \multicolumn{3}{c}{66} \\
\enddata

\tablenotetext{a}{Orbit is consistent with circular, so errors in $\omega$ are large; see
  \citep[see ][ for a fuller explanation.]{Butler06}}
\end{deluxetable}




The residuals to this fit have an RMS 4.03 m/s and show with no
significant periodogram peak at any period.  The tallest peak is
at 3.81 days.  We have run a Monte Carlo FAP analysis on these residuals
of our best fit for this tallest peak, and find similarly good fits in
50\% of velocity-scrambled trials, consistent with noise. We thus conclude that our model is
sufficient to explain the data and that there are no other statistically
significant planetary signals detected.

\section{Newtonian Fits and Stability Analysis}
\label{Newton}
\subsection{MCMC analyses}
We studied the dynamical stability of HD 37124 by combining the radial
velocity data with Markov Chain Monte Carlo (MCMC) analyses to obtain
ensembles of masses, semimajor axes, eccentricities, and orbital
angles consistent with the RV data.  These ensembles were generated
without regard to dynamical stability considerations.  We then imposed
line-of-sight and relative inclination distributions on these sets of
parameters.  By incorporating the unknown inclination parameters with
observation-derived parameters, we sampled the entire phase space of
orbital parameters.   We subsequently ran N-body simulations on each element
in these ensembles in order to assess each system's stability and
resonant evolution.  Our treatment follows that of
\citet{Ford05b,Ford06,Veras09,Veras10}. 

In particular, we calculated 5 Markov chains, each containing over
$10^6$ states.  Each state includes the orbital period ($P$), velocity
amplitude ($K$), eccentricity ($e$), argument of pericenter measured
from the plane of the sky ($\omega$), and mean anomaly at a given
epoch ($u$) for planets b, c and d.   The MCMC uses a standard Gaussian random walk proposal distribution
and the Metropolis-Hastings algorithm for accepting or rejecting each
proposal for all model parameters except $\cos(i_{\mathrm LOS})$ and
$\Omega$.  Since the radial velocity signature is only weakly
dependant on these values, $\cos(i_{\mathrm LOS})$ and $\Omega$ were
drawn randomly for each state.  This can still be considered a Markov
chain, as the procedure satisfies the Markov condition, i.e. that a
trial state not depend on states other than the current state, as well
as the other conditions (time-homogeneous, irreducible, aperiodic) to
prove that the Markov chain will (eventually) converge to the
posterior distribution.

We imposed an isotropic distribution of line-of-sight inclinations ($i_{LOS}$) and a uniform
sample of longitude of ascending nodes ($\Omega$) on our MCMC-derived
initial conditions.  The planet masses, $m$, and semimajor axes, $a$,
were obtained from each set of ($P,K,e,\omega,i,\Omega,u$) values from
relations derived with a Jacobi coordinate system \citep{Lee03}.
The approximate range of minimum masses obtained, in Jupiter masses,
were:  $0.60 \lesssim m_b \sin{i_b} < 0.72$,
$0.40 \lesssim m_c \sin{i_c} < 0.75$, and $0.55 \lesssim m_d \sin{i_d}
< 0.90$. 

We treated the both the offset between the chips and the jitter as free
parameters.\footnote{We adopted a single value of jitter for all
  observations;  in principle the two CCDs may display differing
  amounts of ``instrumental jitter'', such as that due to insufficient
modeling of the charge transfer inefficiencies.  The RMS residuals to
our fit for the two detectors were 3.67 and 4.11 m/s, suggesting that
our assumption of a single jitter value is valid.}  The
5\vth\ percentile, median, and 95\vth\ percentile offsets between the
chips in our ensemble were {3.16, 3.78, 4.62}, and the median jitter
we find to be 4 m/s.

\begin{figure*}
  \centering
  \begin{tabular}{c}
    \multicolumn{1}{c}{\psfig{figure=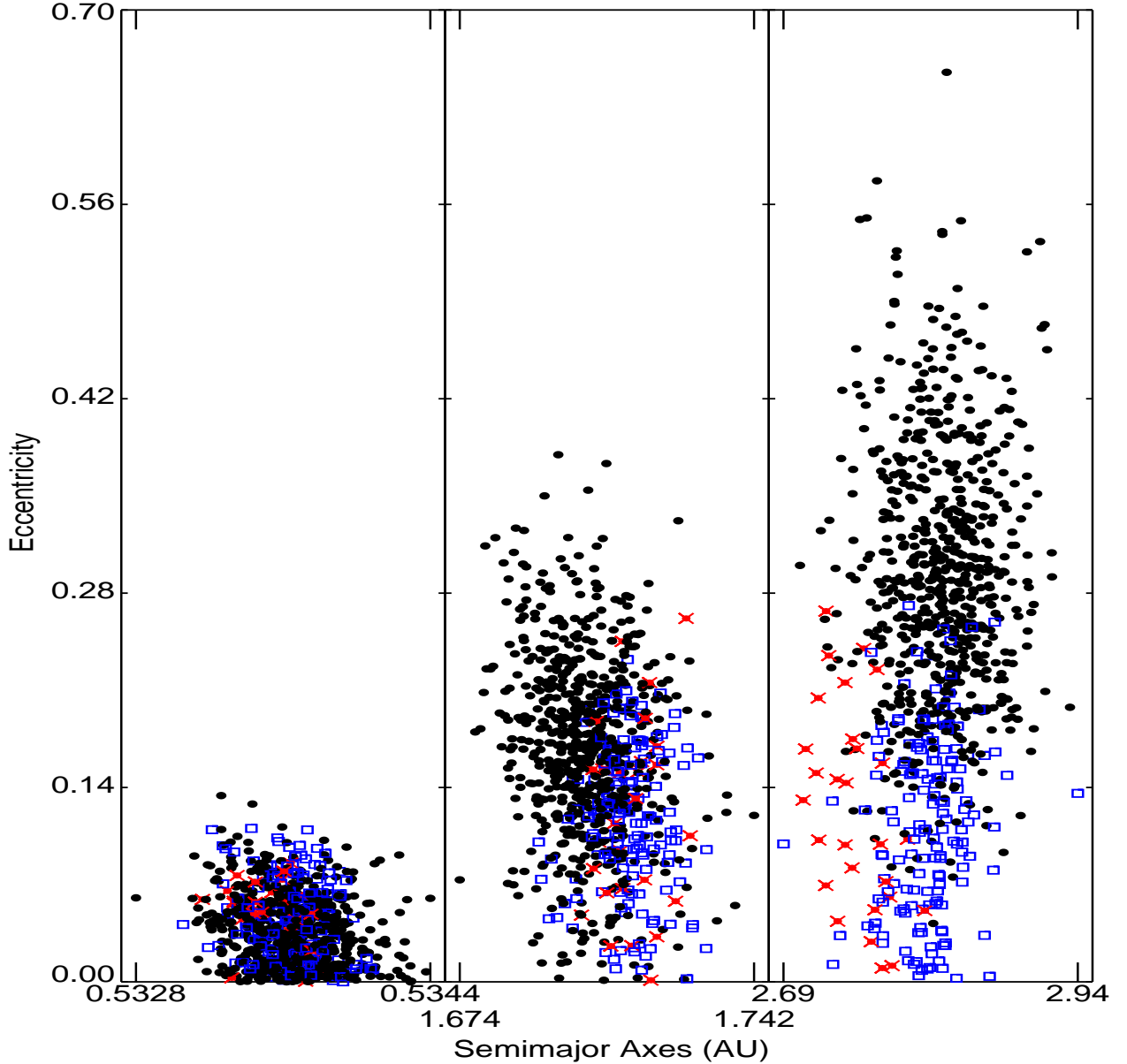,width=1.00\textwidth,height=0.70\textheight}}
\\
  \end{tabular}
\figcaption{Representative eccentricities and semimajor axes of the
  planets in HD 37124.  The three planets are partitioned by panels,
  each with a different horizontal scale.  These ensembles of
  parameters are derived from RV observations using Markov Chain Monte
  Carlo (MCMC) techniques, and represent the initial conditions for a
  subset of our numerical simulations (here, the coplanar prograde
  simulations).  Note that the semimajor axis ratio of the middle and
  outer planets roughly correspond to a $2$:$1$ period commensurability, and
  the inner and middle planets to a $6$:$1$ PC.  Note the changing scale on the $x$-axes in the
  three panels:  the innermost planet is very well constrained in $a$,
  the outermost planet much less so.  The Black dots indicate unstable
  systems, green squares
  represent non-resonant stable systems, and red x's are stable resonant
  systems.  Note that system stability is strongly dependent
  on the eccentricity of the middle and outer planets, and the outer
  planet of resonant systems tend to harbor the smallest initial
  semimajor axes of the ensemble of outer planet ICs. 
\label{mcmc}}
\end{figure*}

\subsection{Coplanar, Prograde Systems}

We integrated 850 sets of initial conditions in the coplanar case with
all three planets in prograde orbits by using the Burlish-Stoer
integrator of {\tt Mercury} \citep{Chambers99} for $10^7$ yr with an
output interval of $10^4$ yr.  We also incorporated the effects of
general relativity in the code, which could have profound consequences
for multi-planet system stability \citep{Veras10}, although the effect is likely to be negligible in this system.  We classified systems as ``unstable'' if, for any planet, $|a_{\rm max}-a_{\rm min}|/a_0 > \tau $, where $a_{\rm max}$, $a_{\rm min}$ and $a_0$ represent the maximum, minimum and initial values of the semimajor axis, and $\tau=0.9$.  

One may visualize a representative architecture of HD 37124 by
comparing the semimajor axis and eccentricity ranges of all three
planets.  Figure \ref{mcmc} plots the observed eccentricity
vs. derived semimajor axis for all planets in the prograde coplanar
state.  Black dots indicate unstable systems while green squares and
red crosses indicate stable systems, and red crosses indicate systems
which are in a $2$:$1$ mean motion resonance (MMR) between planets c
and d, according to our definition below.  The figure indicates i) a
closely packed system, with the inner and outer planets separated by
no more than six times the innermost planet's semimajor axis. ii) a
relatively circular innermost planet (with $e_b \lesssim 0.1$ in most
cases) that is likely too far from the parent star to be classified as
a ``Hot Jupiter'', iii) the greater the number of orbital periods
sampled by RV, the greater the constraint on the planet's likely
semimajor axes and eccentricities, iv) most ($664/850 = 78\%$) current
orbital fits predict an unstable system, v) the majority of initial
conditions which produce stable orbits contain an outer planet with a
low ($< 0.2$) eccentricity, and a middle planet with a semimajor axis
$> 1.695$ AU and eccentricity less than about 0.2., vi) systems
containing a $2$:$1$ resonance occur only when $2.7$ AU $\lesssim a_3
\lesssim 2.8$ AU.  We emphasize that this approximate semimajor axis
range appears to be necessary but not sufficient for resonance to
occur.  The figure demonstrates that other MCMC fits with outer planet
periods in the resonant range are either unstable, or stable but
non-resonant.  The architecture of these systems (as defined by, e.g.,
the mean longitude and longitude of pericenter) do not allow them to
settle into resonance even though the outer planet period might favor
resonance. 

Because of finite sampling, our definition of ``resonance'' in this analysis comes
from consideration of the RMS deviation of each resonant angle about each of $(0^{\circ}, 90^{\circ}, 180^{\circ},
270^{\circ})$, which includes common libration centers.  We flag systems as ``resonant'' if
at least one of these angles has RMS under $90^{\circ}$ for $10$ Myr,
the entire duration of our simulations.  Below, we refer to this value
as a ``libration RMS''.

HD 37124 presents a clear initial choice of angles to test for libration.  As indicated by Fig. \ref{mcmc}, the semimajor axis ratio of planets c and d is suggestive of a $2$:$1$ MMR.  Therefore, we sampled the following angles for libration:

\begin{equation}
\phi_1 \equiv 2 \lambda_d - \lambda_c - \varpi_c  
\end{equation}

\begin{equation}
\phi_2 \equiv 2 \lambda_d - \lambda_c - \varpi_d  
\end{equation}

\noindent{}and found that $\phi_1$ librates in $28/850 = 3.3\%$ of
cases, while $\phi_2$ librates in $9/850 = 1.1\%$ of cases.  Further,
the systems for which $\phi_2$ is resonant are a subset of those for
which $\phi_1$ is resonant.  

\begin{figure*}
 \centering
  \begin{tabular}{ccc}
    \multicolumn{3}{c}{\psfig{figure=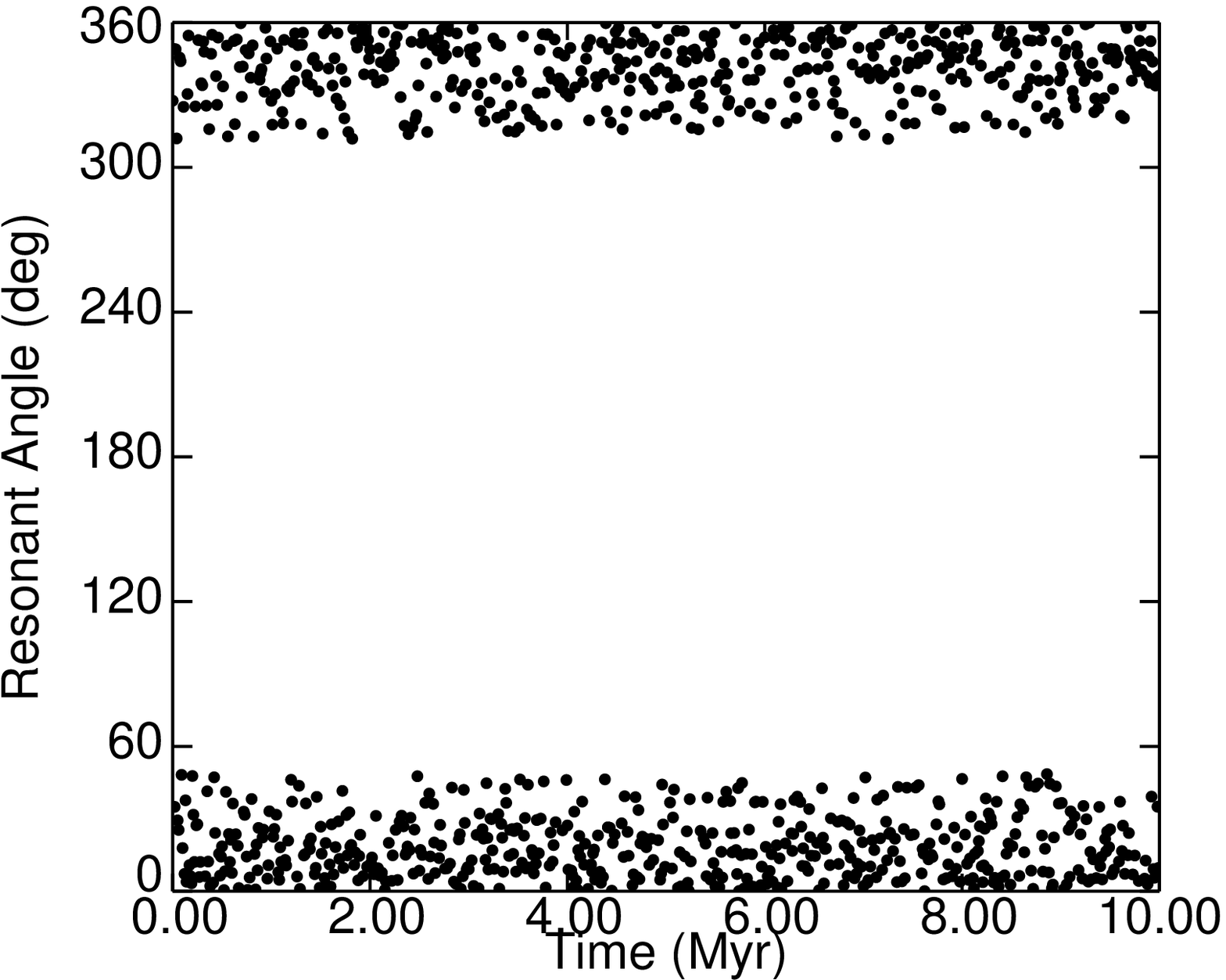,width=0.60\textwidth,height=0.30\textheight}}\\
    \multicolumn{3}{c}{\psfig{figure=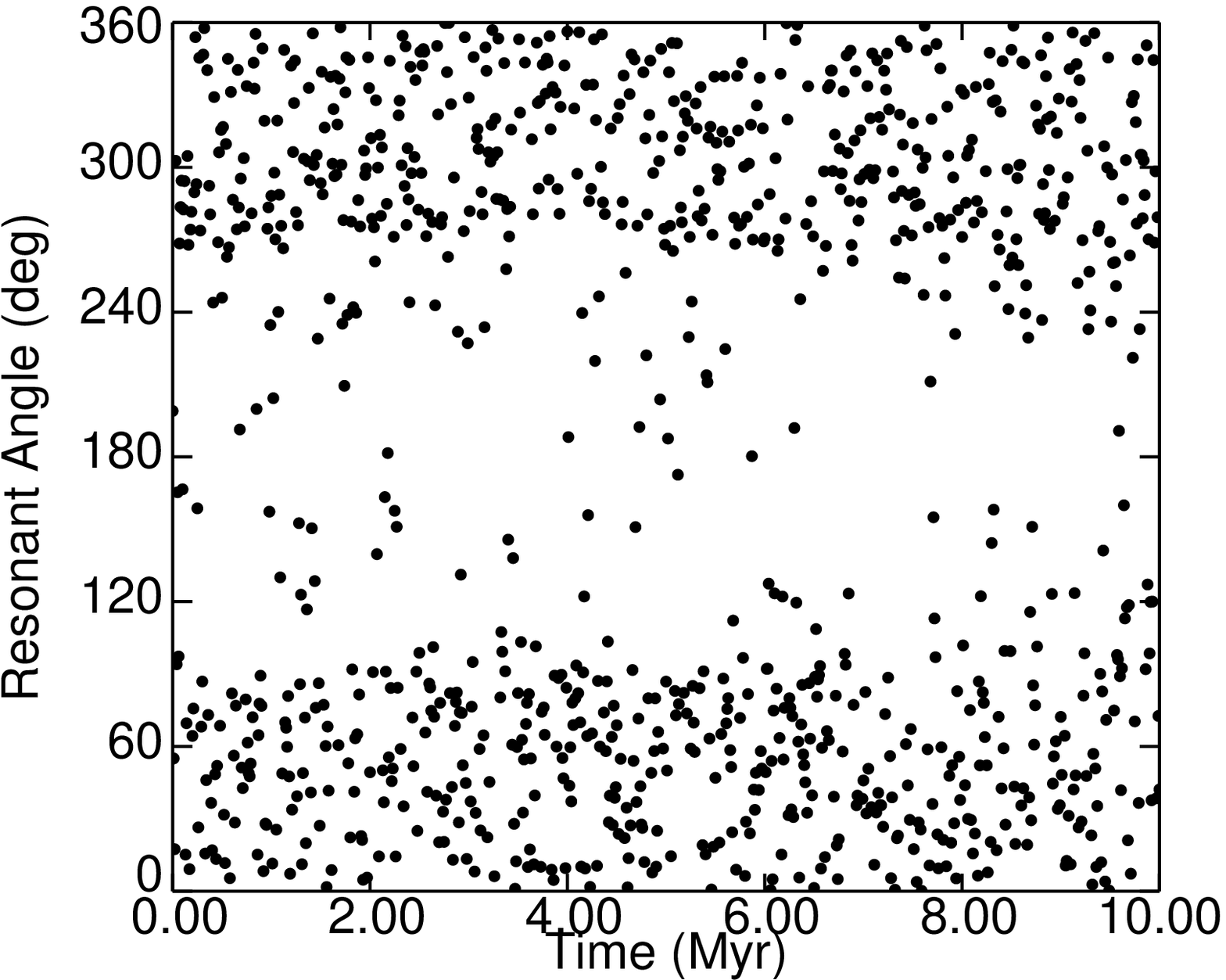,width=0.60\textwidth,height=0.30\textheight}}\\
    \multicolumn{3}{c}{\psfig{figure=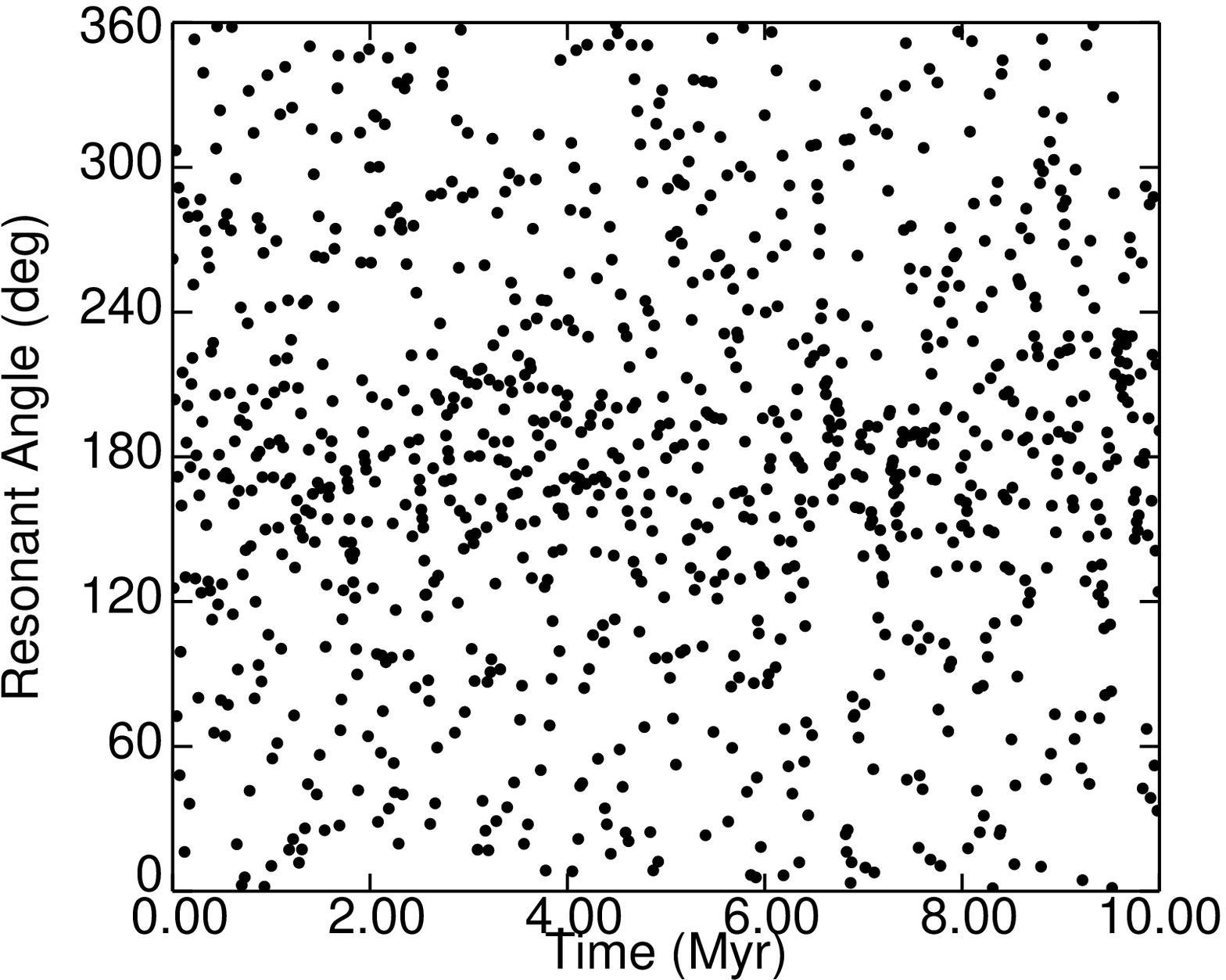,width=0.60\textwidth,height=0.30\textheight}}\\
  \end{tabular}
\figcaption{Three examples of systems that we find to be resonant,
  according to our definition requiring an libration of
  under $90^{\circ}$ for 10 Myr.  Plotted is the time evolution of the
  resonant angle $2 \lambda_d - \lambda_c - \varpi_c$ for a system
  with a computed libration RMS of (upper panel)
  $23.0^{\circ}$ about $0^{\circ}$, (middle panel) $70.2^{\circ}$
  about $0^{\circ}$, and (lower panel) $86.1^{\circ}$ about
  $180^{\circ}$.  
\label{res23}}
\end{figure*}

If we tighten our definition of resonance to include only those
systems with RMS resonant angles under $70^{\circ}$, then no $\phi_2$
arguments are resonant.  Under this stricter definition, the $\phi_1$
arguments are only resonant in $14/850 = 1.6\%$ of cases, and if we
further tighten the libration criterion to an RMS of  or $50^{\circ}$,
then this number decreases to $4/850 = 0.5\%$.  The lowest libration RMS
detected is $23.0^{\circ}$.  All RMS's under
$75^{\circ}$ were for a libration center of $0^{\circ}$.  Figure
\ref{res23} illustrates three examples of ``resonant'' systems from
this, each with a different libration RMS.

\begin{figure*}
  \begin{tabular}{cc}
    \multicolumn{2}{c}{\psfig{figure=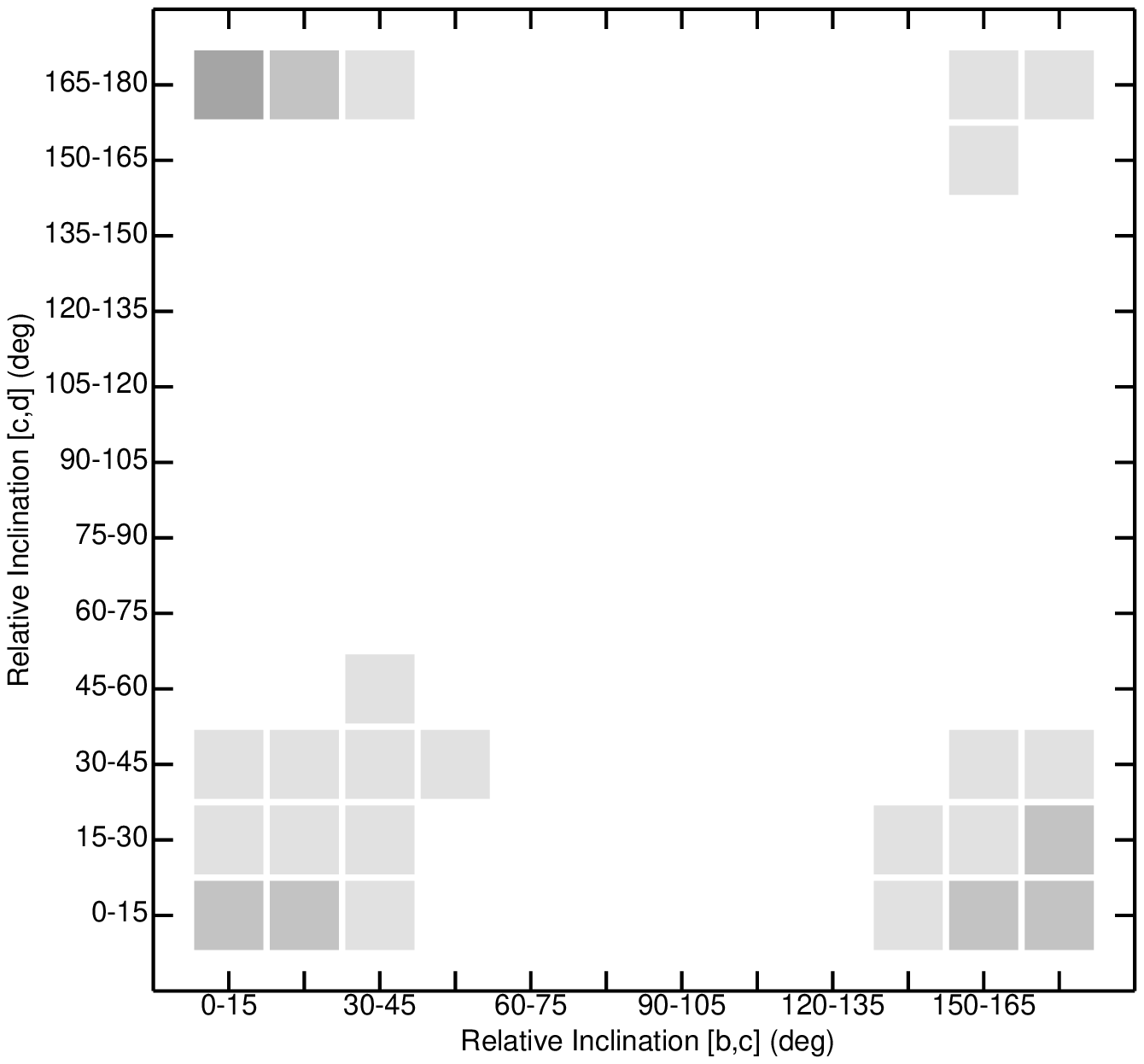,width=0.70\textwidth,height=0.40\textheight}}\\
    \multicolumn{2}{c}{\psfig{figure=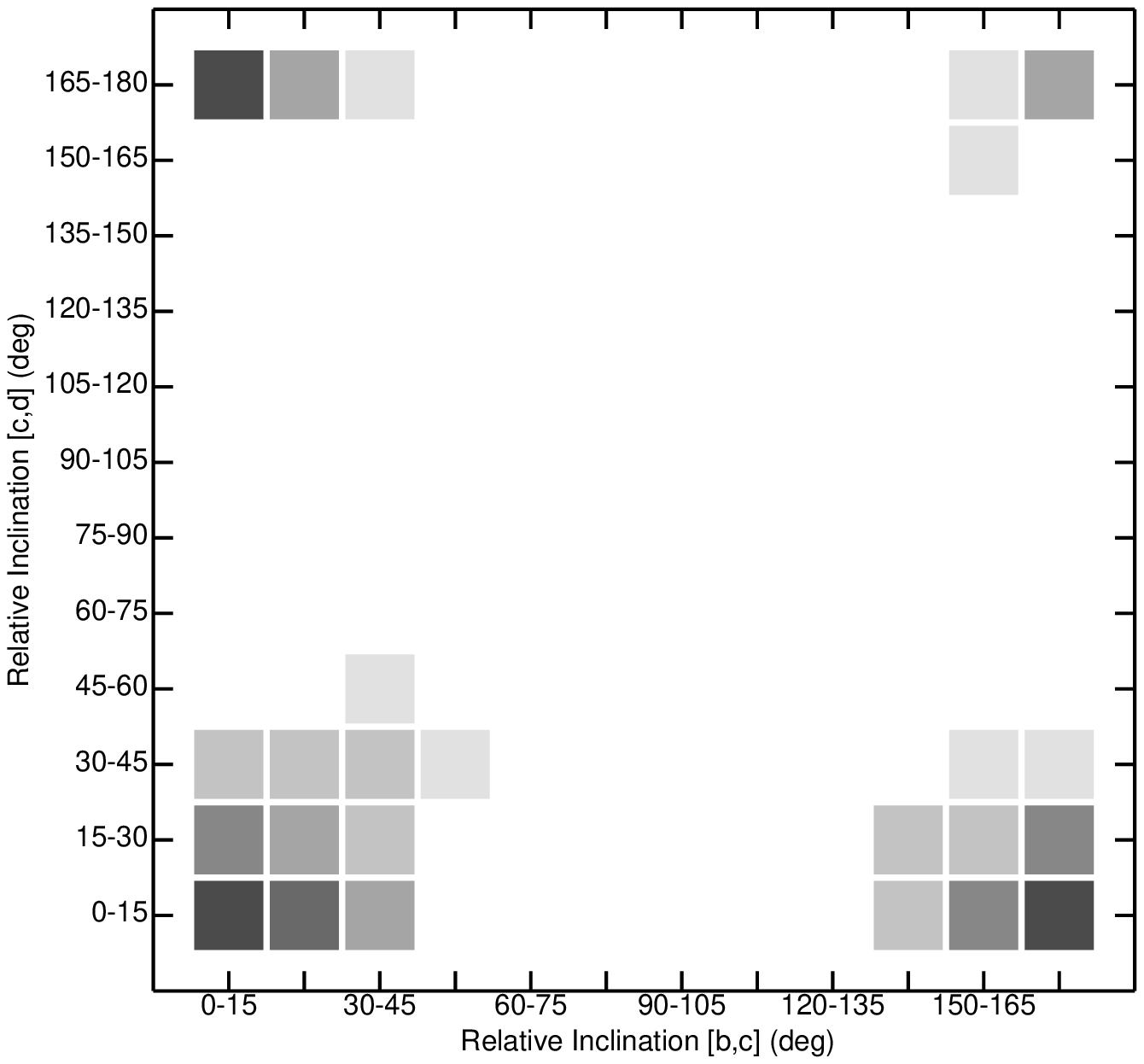,width=0.70\textwidth,height=0.40\textheight}}\\
  \end{tabular}
  \put(-40,-60){\includegraphics[trim = 0mm 0mm 0mm 0mm, clip, width=0.70\textwidth]{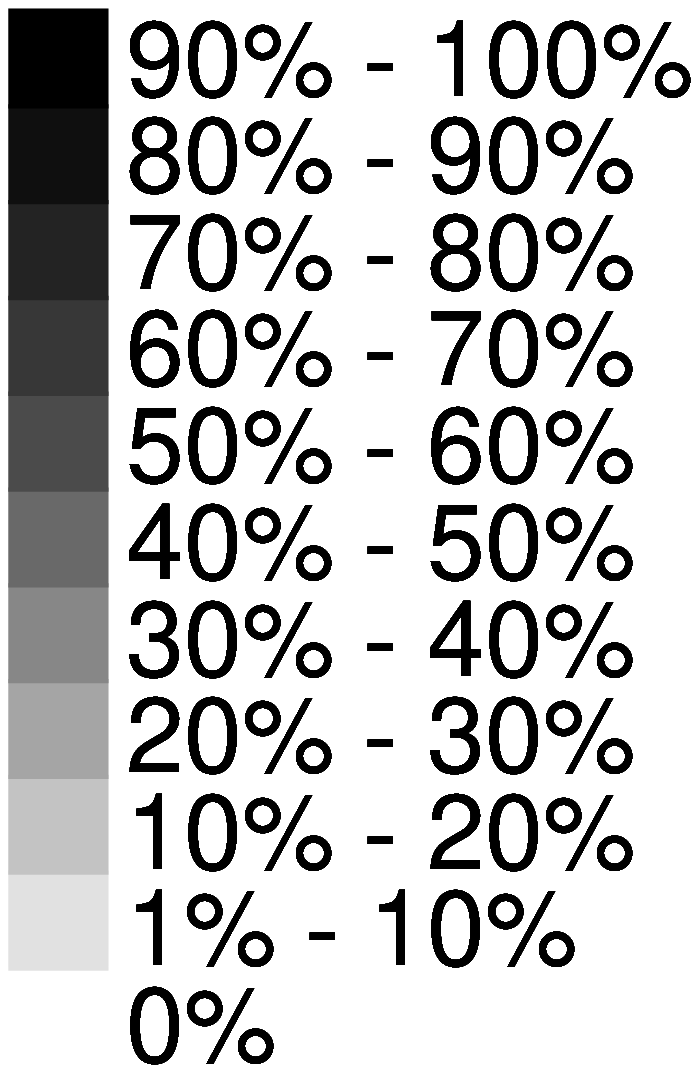}}
\figcaption{Two stability portraits of HD 37124.  Each bin indicates the fraction of stable systems after 10 Myr for all systems (top pallete) and for systems with an initial $e_d < 0.2$ (bottom pallete).  Note that nearly all non-coplanar systems are unstable (indicated by the white spaces), in both various prograde and retrograde cases.  These portraits can provide a useful constraint on the viable relative inclinations in HD 37124. Note, however, that the mutual inclination of planets b and d are not represented on this plot and are only weakly constrained by the other two inclination pairs (e.g. if two pairs are mutually inclined by 30 degrees each, then the third pair may be mutually inclined anywhere between 0 and 60 degrees).
\label{stab}}
\end{figure*}

\begin{figure*}
  \begin{tabular}{cc}
    \multicolumn{2}{c}{\psfig{figure=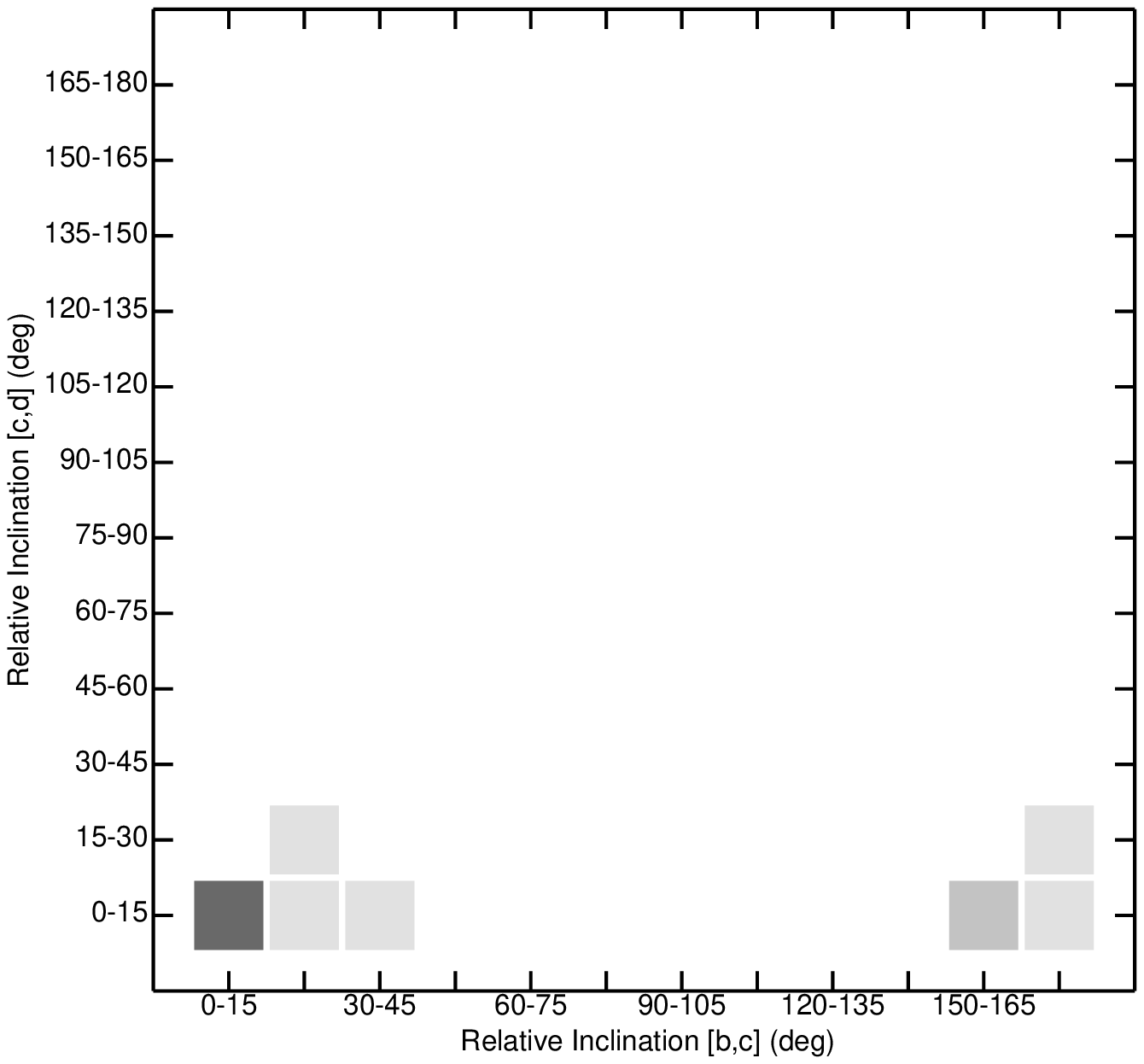,width=0.70\textwidth,height=0.40\textheight}}\\
    \multicolumn{2}{c}{\psfig{figure=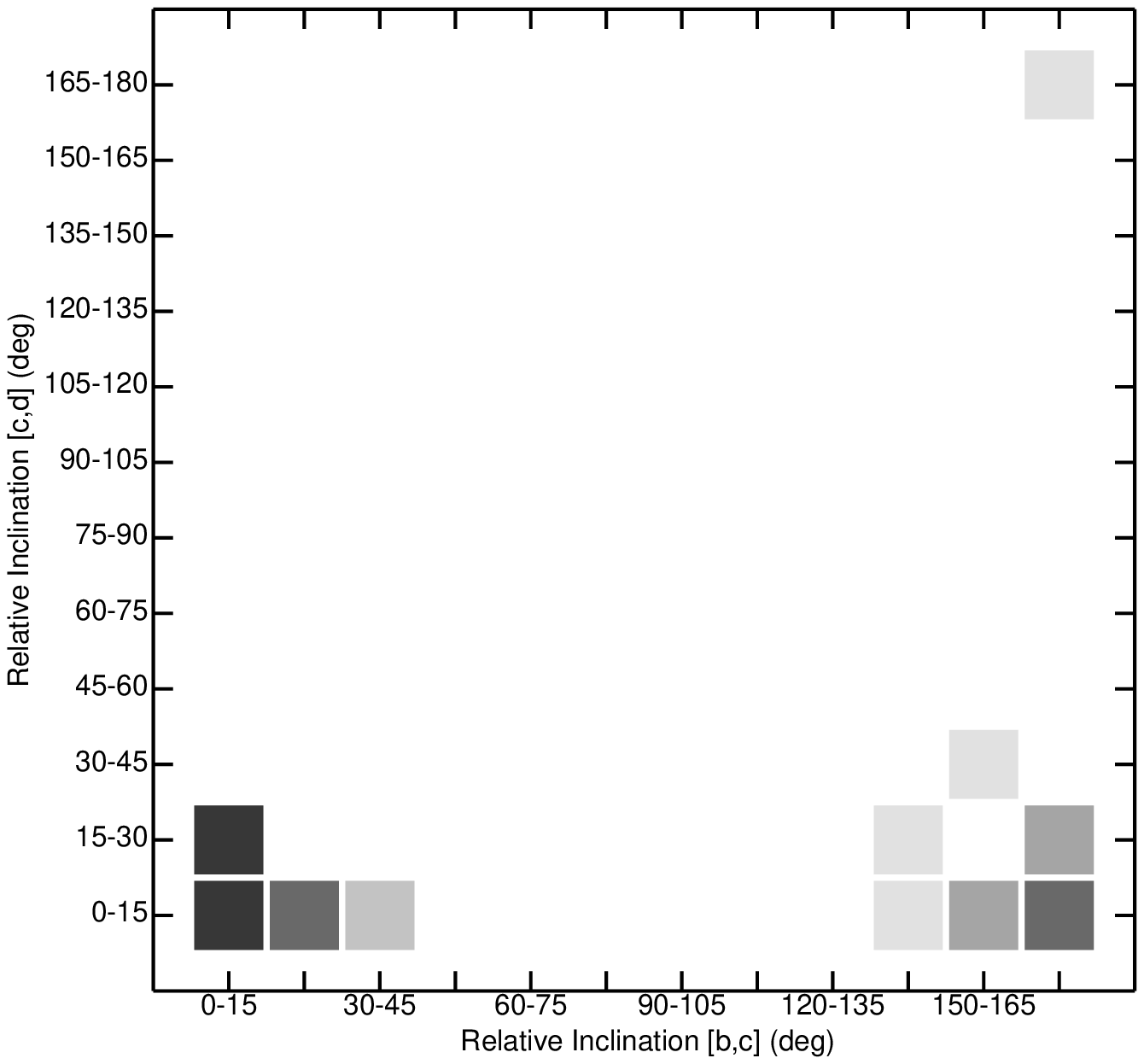,width=0.70\textwidth,height=0.40\textheight}}\\
  \end{tabular}
  \put(-40,-60){\includegraphics[trim = 0mm 0mm 0mm 0mm, clip, width=0.70\textwidth]{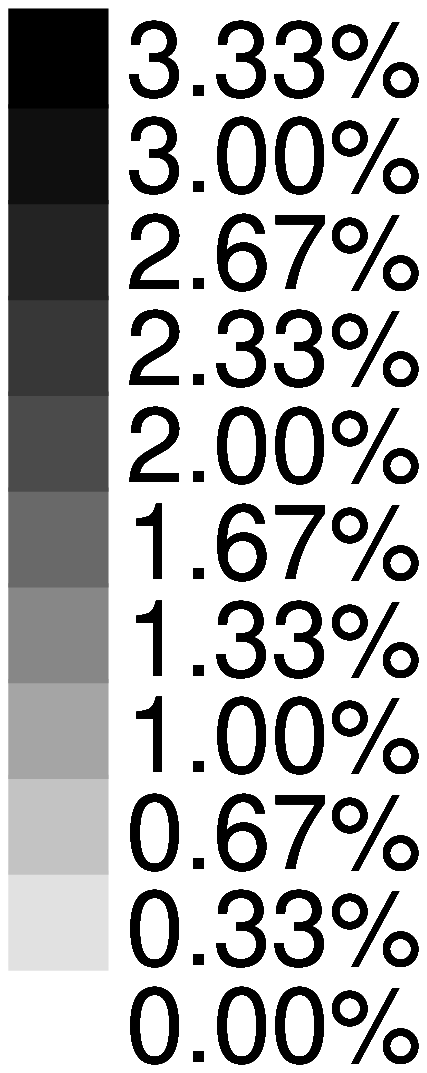}}
\figcaption{Two resonant portraits of HD 37124.  The legend indicates
  the fraction of systems for which the angles $\phi_1 \equiv 2 \lambda_d - \lambda_c - \varpi_c$ (upper panel) and $\phi_1 \equiv 2 \lambda_d - \lambda_c - \varpi_d$ (lower panel) are resonant.  We define ``resonant'' as the situation where the RMS deviation of an angle about a fixed value is less than $90^{\circ}$ over 10 Myr. In most cases, this deviation is between $70^{\circ}$ - $90^{\circ}$, but goes as low as $23^{\circ}$.  Note that only for the near-coplanar cases are any systems resonant.
\label{restables}}
\end{figure*}

We additionally sampled all 3 pairs of apsidal angles (the difference
between two longitudes of pericenter) in the coplanar prograde state,
and found only two instances of libration, both at high ($>
70^{\circ}$) libration RMS's and around the ``asymmetric'' centers
$90^{\circ}$ and $270^{\circ}$ for the inner and outer planet apsidal
angle.  Inspection reveals, however, that these instances of libration
are more indicative of long period ($> 10$ Myr) circulation.

Additionally, the semimajor axis ratio of planets b and c could
indicate the presence of a $6$:$1$ MMR.  Therefore, we sampled all
angles of the form $6 \lambda_d - \lambda_c - t \varpi_c - s
\varpi_d$, where $t+s = 5$.  None of the coplanar prograde systems
exhibited libration of any of the $6$:$1$ angles between planets ``b''
and ``c'' over 10 Myr.  However, preliminary sampling of these angles
over intervals of $2$ Myr does occasionally exhibit libration
RMS's close to $90^{\circ}$.  Because the period ratios between
planets c and d may skirt the $7$:$3$ PC, we also tested the $7
\lambda_d - 3 \lambda_c - t \varpi_c - s \varpi_d$ angles, where $t+s
= 4$, but found no instances of libration. 

\subsection{Mutually Inclined Systems}
Having analyzed the coplanar prograde bin, we can now consider the
case where the planets have nonzero mutual inclinations.  We used
rejection sampling to obtain triplets of $i_{LOS}$ values such that
the system is placed into one of 144 bins according to the relative
inclination between planets $b$ and $c$ ($i_{rel,b,c}$) and planets
$c$ and $d$ ($i_{rel,c,d}$).  In no two bins were the same ensemble of
initial conditions used.  We binned relative inclinations in intervals
of $15^{\circ}$, and used stratified sampling in order to obtain a
uniform number of samples per bin.  We initially sampled 100 initial
system states per bin.  For those bins where we found more than one
system to be stable, we added 200 additional ensembles of initial
conditions

By considering the fraction of stable systems in the non-coplanar
cases, we can obtain a broader dynamical portrait of this system.
Fig. \ref{stab} illustrates the fraction of stable systems in each bin overall (top panel) and with respect to all systems for which the initial $e_d < 0.2$ (bottom panel).  This cutoff was motivated by the rightmost panel in Fig. \ref{mcmc} and could suggest a constraint on the orbital properties of the system in order to ensure that it remains stable.  Fig. \ref{stab} shows that the system {\it must be roughly coplanar}, with relative inclinations less than $\sim 30^{\circ}-45^{\circ}$, in order to be stable.  This constraint allows various pairs of planets to harbor retrograde orbits.  We also performed limited resonant testing for systems in these bins.  The fraction of total systems which exhibit libration of $\phi_1$ and $\phi_2$ under $90^{\circ}$ for 10 Myr is given by Fig. \ref{restables}.  $2$:$1$ resonant systems occur, therefore, generally at the few percent level, and most likely when all planets are coplanar with prograde orbits.

\section{Discussion}

As \citet{Rivera10} showed for the GJ 876 system, even the most
well-established and deepest mean-motion resonances can prove illusory
if additional planets are found in the system (although in that case
it appears that the resonance still present, albiet considerably
shallower and more complex than previously thought).  Even for
truly resonant systems, a demonstration of resonance can be
difficult.  For instance, triple-planet systems may feature two planets with
a mostly-librating resonant argument that occasionally circulates due
to interactions with the third planet.  Near separatrix behavior
\citep[as in the case of $\upsilon$ And;][]{Malhotra02,Ford05} can also
make libration and circulation essentially indistinguishable. 

We note that near-resonant behavior can itself dynamically
interesting: the 5:2 near-resonance of Jupiter and Saturn (the Great
Inequality) has major consequences for the dynamics of the Solar
System.  Given the above-mentioned difficulties in {\it proving} that
a resonant argument for a given system of planets satisfies some precisely
specified definition of libration given the typical uncertainties in
radial velocity measurements, we suggest
that studies of resonant interactions would benefit from identifying 
systems that appear to be in or near resonance ({\it apparent period
  commensuribilities}).  With that in mind, we note that in addition to
HD 37124, there are 21 other systems in the Exoplanet Orbit Database
(Wright et al.\ 2010, PASP submitted) of peer-reviewed\footnote{We have
  included the Kepler multiplanet systems, which had not yet been
  accepted for publication at the time of this writing.} literature
with well-established apparent PCs, which we
present in Table~\ref{mmr}.   This list includes all pairs of planets
for which the period ratio $r$ is less than 6 and within 0.05 of an integer or half
integer (neglecting uncertainties in periods), and other
exoplanetary pairs whose likely MMRs are discussed in 
the literature. 

\begin{deluxetable*}{lccl}
\tablecolumns{4}
\tablecaption{Apparent period commensuribilities in well-characterized
  multiplanet systems\label{mmr}}
\tablehead{\colhead{System} & \colhead{Components} & \colhead{Period Ratio} &
  \colhead{References}}
\startdata
GJ 876 & $e,c,b$ & 4:2:1 & \citet{Marcy_gl876,Rivera10}\\
HD 82943 & $b,c$ & 2:1 & \citet{Mayor04,Ji03,Lee06}\\ 
HD 37124 & $c,d$ & 2:1 & \citet{Vogt05}, this work \\
HD 128311 & $c,b$ & 2:1 & \citet{SandorScatter,Vogt05}, others \\
HD 73526 & $c,b$ & 2:1 & \citet{Tinney06,Sandor06}\\ 
$\mu$ Ara & $b,e$ & 2:1 & \citet{Pepe07,Gozdziewski07}\\
KOI 152\tablenotemark{$\dagger$} & 2,3 & 2:1 & \citet{Steffen10} \\ 
KOI 877\tablenotemark{$\dagger$} & 2,1 & 2:1 & \citet{Steffen10} \\ 
24 Sex & $c,b$ & 2:1 & \citet{Johnson11} \\
Kepler-9 & $c,b$ & 2:1 & \citet{Holman10} \\
PSR B1257+12 & $B,C$ & 3:2 & \citet{Wolszczan92,Malhotra92,Konacki99b} \\  
HD 45364 & $c,b$ & 3:2 & \citet{Rein10,Correia09}\\ 
HD 200964 & $c,b$ & 4:3 & \citet{Johnson11}\\
55 Cnc & $c,b$ & 3:1 & \citet{Fischer08,Zhou08} \\  
HD 10180& $d,e$ & 3:1 &Lovis et al.\ 2010, A\&A accepted \\  
HD 60532 & $c,b$ & 3:1 & \citet{Desort08,Desort09,Laskar09}\\
HD 108874 & $c,b$ & 4:1 & \citet{Vogt05,Gozdziewski06b}\\  
Solar & $\saturn,\jupiter$ & 5:2 & \\
HD 10180& $e,f$ & 5:2 & Lovis et al.\ 2010, A\&A accepted\\ 
KOI 896\tablenotemark{$\dagger$} & 1,2 & 5:2 & \citet{Steffen10} \\
HD 202206 & $c,b$ & 5:1 & \citet{Correia05,Gozdziewski06b}  \\
\enddata
\tablenotetext{$\dagger$}{This is a candidate exoplanet system based on
  {\it Kepler} photometry, but the planets are at present considered
  unconfirmed.  For example, KOI 877 could, in principle, be a blend
  of two stars, each hosting one transiting planet in a coincidental
  apparent PC.}
\end{deluxetable*}

The fraction of known multiplanet systems exhibiting at least one
apparent PC is high.  Of the 43 well determined multiplanet systems
discovered by radial velocities around normal stars, 15 appear in Table~\ref{mmr}, or 35\%, including 9 of the 
30 apparent double-planet systems 30\%.\footnote{We have excluded in this statistic the Kepler
 systems, the pulsar system, the microlensing system, planets from
  direct imaging, and the Solar System.  We acknowledge that a more rigorous
  statistic would be valuable, but note that it would need to address some strong 
  detectability and selection effects regarding planets in multiple
  planet systems, and to assess these detection thresholds across multiple,
  heterogeneous surveys.  To give just one example, we note that in
  addition to a radial velocity survey's decreasing sensitivity to planets in longer periods,
  it can be difficult to detect an interior planet in a 2:1 resonance
  due to approximate degeneracy with eccentricity in a single planet
  model \citep[e.g.,][]{Anglada-Escude10,Moorhead10}.  Such an analysis is beyond
  the scope of this manuscript.}   To determine if this is more than
would be expected simply by chance, we have performed two tests.  

In the first test, we randomly drew pairs of periods from the 340
RV-discovered planets in the Exoplanet Orbit Database (EOD)
\citep[][PASP submitted]{Wright10}, and rejected those pairs with
period ratios $r < 1.3$ (corresponding to the smallest $r$ among real
multiplanet systems).  We counted the fraction of remaining systems
with $r$ within 0.05 of an integer or half integer $\leq 5$
(corresponding to the largest apparent PC in Table~\ref{mmr}).  We
found that only 4\% of our random pairs satisfy our apparent PC criterion, far  
smaller than the 30\% of double-planet systems actually found in
apparent PCs.

In the second test, we included the effects of triple and higher-multiple
systems by randomly assigning periods from the EOD to all planets in
real multiple systems (again subject to the constraint that no pair of
planets in the system have $r < 1.3$).  We found 16\% of these
artificial systems passed our apparent PC criterion, reflecting the
higher number of planet pairs available to test per star compared 
to our first test.  Despite this inflation, the actual value of 33\%
among all multiplanet systems is still significantly
higher. \footnote{We suspect that the reason the observed value is not similarly inflated with
  respect to double planet systems is that our randomization did not include the requirement of dynamical
  stability, as real systems implicitly do.}  These results underscore that the orbital periods of the
population of planets known to be in multiplanet systems is
inconsistent with the apparently-singleton sample \citep{Wright09}.

This apparently high percentage of known multi-planet systems with an apparent PC
might  favor particular formation mechanisms.  Planet-planet scattering, planetesimal disk migration, and
gas disk migration have all been shown to produce systems with at least one pair of planets
that not only are commensurate in period, but also resonant.
\citet{Raymond08} found that planet-planet
scattering produced MMRs in roughly 5\% of the systems that they simulated, and \citet{Raymond10}
discovered that between 50\% - 80\% of systems undergoing planetesimal disk migration yielded
resonant capture.  Convergent gas disk migration, the thrust of the numerous papers cited in \S\ref{Intro}
can occur with near 100\% efficiency for certain initial planetary and disk parameters.
As observed by \citet{Thommes03} and \citet{Libert09}, the inclination may be excited
as well as the eccentricity in many resonant cases.  If a resonance exists in HD 37124, it could
have been produced by any of these methods.   If the RMS libration of
such a resonance is representative of the bottom panel of
Fig.~\ref{res23} and has a value that approaches 90 degrees, then
planet-planet scattering is a likely origin of this resonance.
Alternatively, disk or gas migration would likely produce a system
that is "deeper" in resonance, with a smaller variation in resonant
angle, similar to the top panel of Fig.~\ref{res23}.  
 Resonant librating angles
need not involve the eccentricities and pericenters, but instead the inclinations and longitudes of ascending nodes,
similar to the $4$:$2$ Mimas-Tethys resonance in the Saturnian system
\citep{Champenois99a,Champenois99b}.  

\section{Conclusions}

We have resolved the period ambiguity of HD 37124 $d$ from
\citet{Vogt05} and find that HD 37124 $c$ and $d$ are in an 
apparent 2:1 period commensurability.  Our numerical integrations show
that both resonant and non-resonant configurations are consistent with
the radial velocity data, and that stability requires a nearly
circular orbit ($e < 0.3$) for the $d$ component.  Our stability
analysis shows that the system must be
nearly coplanar, and that the three planets have identical minimum masses
within the errors (of 3--10\%). 

We show that the roughly one in three well-characterized multiplanet
systems shows an apparent period commensurability, which is more than
a na\"ive estimate based on randomly drawing periods from the known
exoplanet population would suggest.  This offers evidence for some
particular proposed scattering and migration mechanisms, and we
suggest that the statistics of multiplanet systems may now be
sufficiently robust to provide a test and comparison of models
of exoplanetary dynamical evolution.

\acknowledgments

We thank the referee, Daniel Fabrycky, for his constructive and thorough
review, which siginifacntly improved this manuscript.

J.T.W.\ received support from NASA Origins of Solar Systems grant
NNX10AI52G.  D.V. and E.B.F. were partially supported by NASA Origins of Solar Systems grant
NNX09AB35G.   A.W.H.\ gratefully acknowledges support from a Townes Postdoctoral Fellowship at the UC Berkeley Space Sciences Laboratory.  

This work was partially supported by funding from the Center for Exoplanets
and Habitable Worlds, which is supported by the Pennsylvania State University, the Eberly College
of Science, and the Pennsylvania Space Grant Consortium.  

The work herein is based on observations obtained at the W. M. Keck
Observatory, which is operated jointly by the University of California
and the California Institute of Technology.  The Keck Observatory was
made possible by the generous financial support of the W.M. Keck
Foundation.  We wish to recognize and acknowledge the very significant
cultural role and reverence that the summit of Mauna Kea has always
had within the indigenous Hawaiian community.  We are most fortunate
to have the opportunity to conduct observations from this mountain.

The authors acknowledge the University of Florida High-Performance
Computing Center for providing computational resources and support
that have contributed to the results reported within this paper.  

This research has made use of NASA's Astrophysics Data System and the
Exoplanet Orbit Database at exoplanets.org.

\facility{{\it Facility} Keck:I}

\end{document}